\documentclass[preprint,12pt,authoryear,times]{elsarticle}
\usepackage[T1]{fontenc}
\usepackage[utf8]{inputenc}
\usepackage{courier}
\usepackage{amsmath,amssymb,amsthm,mathtools}
\usepackage{booktabs}
\usepackage{array}
\usepackage{multirow}
\usepackage{graphicx}
\usepackage{xurl}
\usepackage{hyperref}
\usepackage{xcolor}
\hypersetup{colorlinks=true,citecolor=blue,linkcolor=blue,urlcolor=blue}

\newcommand{\bp}{\ensuremath{\,\mathrm{bp}}}
\newcommand{\pair}[1]{\ensuremath{\text{\normalfont #1}}}
\newtheorem{proposition}{Proposition}

\begin{document}

\begin{frontmatter}

\title{Ethereum as a Relative Asset-Pricing Problem Across Numeraires: Baseline Evidence from Triangular Pricing Consistency}

\author[olin]{Hongzhe Wen\textsuperscript{*}}

\affiliation[olin]{
  organization={Olin Business School, Washington University in St. Louis},
  city={St. Louis},
  state={MO},
  country={United States}
}

\begin{abstract}
This paper studies Ethereum pricing as a relative asset-pricing problem across numeraires. The core restriction is triangular consistency among \pair{ETH/USD}, \pair{ETH/BTC}, and \pair{BTC/USD}: the direct \pair{ETH/USD} price should equal the synthetic price implied by the other two legs. Using a single-venue Coinbase true-USD triangle at hourly frequency from January 2022 through June 2023, I measure the log wedge between direct and synthetic prices. The wedge is close to zero on average, ADF diagnostics reject the unit-root null, and first-order persistence is low, while its absolute magnitude rises during stress episodes, especially around the FTX collapse. Supporting evidence from a Binance USDT triangle shows a tighter unconditional wedge, suggesting that venue liquidity and arbitrage efficiency also matter for wedge magnitudes in this sample. A Gate.io order-book extension for the Terra/Luna, FTX, and USDC--SVB windows uses hourly bid and ask snapshots to construct executable triangular round trips; across 524 common event-hour observations, no positive after-fee executable breach is observed. The results support a conservative interpretation: Ethereum prices generally align closely with triangular consistency, while stress-state deviations are best understood as relative pricing wedges within market frictions rather than as persistent intrinsic misvaluation.
\end{abstract}


\end{frontmatter}

\vfill
\begingroup
\footnotesize
\noindent\rule{0.38\textwidth}{0.4pt}\par
\noindent\textsuperscript{*}Corresponding author. Email address: \href{mailto:hongzhewen01@gmail.com}{hongzhewen01@gmail.com} (Hongzhe Wen).
\endgroup
\clearpage

\section{Introduction}
Many applied discussions of Ethereum begin with a valuation question: what determines the level of ETH priced in dollars, and which fundamentals should anchor that dollar price. This paper starts from a different question. For an asset that is simultaneously traded against USD, BTC, and other crypto-market benchmarks, one useful empirical question is whether Ethereum is priced consistently across numeraires. In that sense, the relevant object is not a dollar-denominated intrinsic-value gap, but a relative pricing restriction written across directly traded spot rates.

This paper develops that idea in a deliberately conservative way. The core restriction is the triangular identity linking \pair{ETH/USD}, \pair{ETH/BTC}, and \pair{BTC/USD}. If the direct and synthetic \pair{ETH/USD} prices coincide, then Ethereum pricing is cross-numeraire consistent at that instant. If they do not coincide, the discrepancy is measured by a relative pricing wedge,
\[
\epsilon_t = \log P_t^{\pair{ETH/USD}} - \log P_t^{\pair{ETH/BTC}} - \log P_t^{\pair{BTC/USD}}.
\]
The object \(\epsilon_t\) is interpreted throughout as a parity deviation or triangular consistency wedge. It is not interpreted as evidence that Ethereum is intrinsically overvalued or undervalued in dollars.

Framed this way, the paper sits closer to the literature on no-arbitrage restrictions, change of numeraire, triangular parity, and limits to arbitrage than to the literature proposing ever-richer valuation formulas for ETH. The formal starting point is standard: change-of-numeraire arguments write pricing problems in relative units rather than treating one asset as a uniquely privileged anchor, as in the measure-change framework of \citet{geman_elkaroui_rochet_1995}. \citet{cochrane_2005} provides the broader asset-pricing language for distinguishing a pricing restriction from a structural valuation model.

The empirical complication is equally standard: an observed parity deviation is not automatically a costless arbitrage. \citet{fenn_etal_2009} study triangular arbitrage using high-frequency executable FX prices, while \citet{akram_rime_sarno_2008} show in related real-time FX arbitrage evidence why synchronization and execution matter. \citet{chaboud_etal_2014} shows how algorithmic trading changes the frequency of triangular-arbitrage opportunities. In crypto markets, segmentation and uneven arbitrage capacity make those issues at least as relevant \citep{makarov_schoar_2020}.

The baseline empirical contribution uses a deliberately conservative design: a single-venue Coinbase spot triangle with true USD pairs, hourly observations, intersection-only synchronization, and no forward fill. This design intentionally narrows the scope of interpretation. Because all three legs are native Coinbase spot markets and the dollar leg is true USD rather than a stablecoin proxy, the baseline triangle is a narrow reduced-form measurement of cross-numeraire consistency within one venue. It does not identify stablecoin-specific mechanisms, and it is not intended to.

Within that baseline, three findings are central. First, the wedge has a near-zero mean. Over 13{,}100 hourly observations from 2022-01-01 to 2023-06-30, the sample mean is approximately \(-0.065\bp\), which is economically indistinguishable from zero. Second, the wedge shows little persistence rather than a slow drift. The diagnostics report an Augmented Dickey--Fuller statistic of \(-110.893\), continued unit-root rejection under lag augmentation, an AR(1) coefficient of approximately \(0.0315\), and an implied half-life of about 0.20 hours. Third, parity deviations widen during stress episodes. Mean absolute wedges are larger during the Terra/Luna, FTX, and USDC--SVB windows than in calm periods, with the FTX episode showing the strongest event-study contrast by the Welch statistic.

Three completed extensions sharpen, rather than overturn, that baseline interpretation. The first is a Binance spot triangle that uses USDT as the dollar proxy over the same sample window. In unconditional terms, that extension produces a materially tighter wedge than the Coinbase true-USD baseline. The second is a Gate.io order-book extension over the Terra/Luna, FTX, and USDC--SVB stress windows. It uses genuine bid and ask snapshots for \pair{ETH/USDT}, \pair{ETH/BTC}, and \pair{BTC/USDT}, and finds that the event-hour midquote wedges remain inside bid--ask plus fee costs. These extensions do not support treating stablecoin denomination as the dominant average source of pricing noise in this sample; they instead point to venue-level liquidity, quote depth, and arbitrage efficiency as additional determinants of wedge magnitudes. Stablecoin-related mechanisms remain relevant as a stress-state hypothesis, but the completed extensions counsel against treating them as the leading unconditional explanation. The third extension attempted to move the Coinbase baseline from close-based measurement to a quote-based interpretation. However, no historical top-of-book midquote series is available for the Coinbase true-USD sample; the resulting high--low midrange proxy is not a valid substitute for synchronized historical midquotes in a triangular pricing test. The paper therefore reports that exercise as a methodological limitation and negative result rather than as a robustness success.

These results support a modest interpretation. They are consistent with the view that triangular pricing holds tightly on average in normal conditions, but that the wedge widens when market stress raises frictions or changes the speed with which relative prices adjust. That is closer to the language of parity wedges in constrained arbitrage settings \citep{du_tepper_verdelhan_2018} and to the broader limits-to-arbitrage logic of \citet{shleifer_vishny_1997} than to any claim of persistent intrinsic mispricing. The paper therefore treats nonzero \(\epsilon_t\) as a state-dependent parity deviation, not as a direct measure of fundamental misvaluation.

The paper's contribution is correspondingly narrow and deliberate. It does not propose a new structural valuation model for Ethereum. Instead, it argues that Ethereum research can impose a relative asset-pricing consistency benchmark across numeraires before asking which mechanisms explain departures from that benchmark. The headline evidence is restricted to the Coinbase true-USD triangle. Stablecoin-proxy and executable-band evidence are reported as supporting extensions, while higher-frequency quote reconstruction and broader cross-venue mechanism tests define the next research margin.

The rest of the paper proceeds as follows. Section~\ref{sec:related} places the paper in the literatures on triangular parity, numeraire choice, parity wedges, and crypto market structure. Section~\ref{sec:theory} presents the triangular pricing restriction and defines the wedge. Section~\ref{sec:data} describes the baseline data and variable construction. Section~\ref{sec:design} outlines the empirical design and distinguishes the headline baseline from supporting extensions. Section~\ref{sec:baseline} reports the baseline evidence. Section~\ref{sec:stress} discusses stress episodes. Section~\ref{sec:discussion} interprets the findings conservatively. Section~\ref{sec:conclusion} concludes.

\section{Related Literature}\label{sec:related}
This paper speaks to several related literatures.

First, it draws on the triangular-arbitrage and triangular-pricing literature in foreign exchange. \citet{fenn_etal_2009} show that triangular-arbitrage opportunities in spot FX are short-lived, small, and measured most meaningfully with high-frequency executable prices. \citet{akram_rime_sarno_2008} provide related real-time FX evidence showing that apparent arbitrage opportunities depend on synchronization, execution, and market microstructure. \citet{cui_etal_2020} formalize triangular-arbitrage detection under nontrivial transaction costs. \citet{gradojevic_erdemlioglu_gencay_2020} show that limit-order-book conditions, realized volatility, and cross-correlations help predict ultra-high-frequency triangular arbitrage opportunities.

The same literature also disciplines how this paper interprets nonzero wedges. \citet{chaboud_etal_2014} show that algorithmic trading changes the frequency of triangular-arbitrage opportunities in FX, and \citet{ito_yamada_takayasu_2020} emphasize execution risk when a trader must complete multiple legs. \citet{reynolds_sogner_wagner_2021} provide a close bridge to crypto-relevant triangular parity analysis by studying deviations involving Bitcoin. This literature motivates the paper's emphasis on synchronized pricing and on the distinction between a measured wedge and a realizable arbitrage profit.

Second, the paper is informed by the change-of-numeraire tradition in asset pricing. \citet{geman_elkaroui_rochet_1995} formalize the link between numeraire changes and probability-measure changes in pricing problems. \citet{schroder_1999} uses the same logic to derive parity results for futures, forwards, and options. The point used here is narrower: relative pricing relations can be written across numeraires without treating the dollar leg as the only natural anchor. \citet{cochrane_2005} is especially useful because it clarifies the distinction between a general pricing restriction and a structural model of cash-flow-based valuation. That distinction is central for Ethereum, where much of the applied discussion begins with a dollar valuation question rather than a consistency question.

Third, the paper is related to work on parity wedges and constrained arbitrage in international finance. \citet{du_tepper_verdelhan_2018} document large and persistent covered-interest-parity deviations after the global financial crisis. \citet{cerutti_obstfeld_zhou_2021} study macrofinancial determinants of those deviations. This tradition supplies the analogy for the present design: a crypto triangular wedge can be studied as an economically meaningful pricing object even when it does not imply frictionless arbitrage. This perspective is reinforced by the broader limits-to-arbitrage framework of \citet{shleifer_vishny_1997}.

Fourth, the paper connects to the literature on crypto market structure and segmentation. \citet{makarov_schoar_2020} document large and recurrent cross-market cryptocurrency price deviations. More recent evidence suggests that these opportunities have become harder to exploit as market efficiency has improved \citep{crepelliere_pelster_zeisberger_2023}, and that the introduction of Bitcoin futures improved spot-market synchronicity \citep{augustin_rubtsov_shin_2023}.

This market-structure literature is useful because it gives a mechanism, not just background. \citet{saggese_etal_2023} provide micro-level evidence that crypto arbitrage activity can be concentrated among a small set of sophisticated traders. \citet{shynkevich_2023} directly compares law-of-one-price deviations and arbitrage trading in Bitcoin and Ethereum. These papers are not tests of the Coinbase triangle studied here, but they motivate the idea that small average deviations can coexist with stress-state widening when arbitrage capacity is uneven.

Fifth, the paper is deliberately distinct from both general crypto asset-pricing models and Ethereum protocol-economics work. The return-factor literature studies expected returns and common risk exposures in cryptocurrency markets; \citet{liu_tsyvinski_2021} is the benchmark example. Ethereum-focused mechanism-design papers ask a different set of questions: \citet{roughgarden_2020} studies EIP-1559's transaction-fee design, while \citet{liu_etal_2022_eip1559} evaluates its empirical effects on fees, waiting times, and consensus security. Those studies ask what determines returns, network-resource prices, or protocol-level incentives. By contrast, the present paper asks whether observed spot prices of ETH are mutually consistent across numeraires. In that sense, the contribution is complementary to valuation and protocol-economics work, not a substitute for it.

A final adjacent literature concerns stablecoin stress and dollar-proxy fragility. \citet{ahmed_aldasoro_duley_2024} show how public information can affect stablecoin run dynamics, and \citet{ma_zeng_zhang_2025} emphasize the centralization of arbitrage in stablecoin runs. \citet{galati_capalbo_2024} study the USDC--SVB episode as a shock to stablecoin stability. \citet{badev_watsky_2023} study DeFi spillovers from the Terra collapse. That channel matters for interpretation, but it is not the baseline mechanism because the Coinbase triangle excludes USDT and other stablecoin dollar proxies by construction.

\section{Theoretical Framework}\label{sec:theory}
\subsection{Triangular pricing consistency}
Let \(P_t^{\pair{ETH/USD}}\) denote the direct spot price of ETH in USD, \(P_t^{\pair{ETH/BTC}}\) the spot price of ETH in BTC, and \(P_t^{\pair{BTC/USD}}\) the spot price of BTC in USD. In frictionless and synchronized trading, internal consistency requires
\begin{equation}
P_t^{\pair{ETH/USD}} = P_t^{\pair{ETH/BTC}} \times P_t^{\pair{BTC/USD}}.
\label{eq:triangle-level}
\end{equation}
Equation \eqref{eq:triangle-level} is not a new valuation formula. It is a relative pricing restriction implied by internal consistency across directly traded spot rates.

Taking logs yields the additive representation
\begin{equation}
\log P_t^{\pair{ETH/USD}} = \log P_t^{\pair{ETH/BTC}} + \log P_t^{\pair{BTC/USD}}.
\label{eq:triangle-log}
\end{equation}
The paper measures departures from \eqref{eq:triangle-log} with
\begin{equation}
\epsilon_t \equiv \log P_t^{\pair{ETH/USD}} - \log P_t^{\pair{ETH/BTC}} - \log P_t^{\pair{BTC/USD}}.
\label{eq:epsilon}
\end{equation}
When \(\epsilon_t=0\), the triangle is exactly consistent. When \(\epsilon_t\neq 0\), the direct and synthetic \pair{ETH/USD} prices differ.

\subsection{Interpretation of the wedge}
The wedge \(\epsilon_t\) is interpreted as a relative pricing wedge, parity deviation, or deviation from triangular consistency. This terminology matters. A nonzero \(\epsilon_t\) need not imply an exploitable round-trip arbitrage after bid--ask spreads, fees, timing risk, algorithmic competition, and inventory constraints are taken into account. \citet{fenn_etal_2009} make this distinction in triangular FX evidence based on high-frequency executable prices, while \citet{akram_rime_sarno_2008} make a related point for real-time FX arbitrage evidence more generally. The transaction-cost formalization in \citet{cui_etal_2020} makes the same point from a detection perspective. Nor does \(\epsilon_t\) identify a latent intrinsic value of Ethereum. It instead measures the extent to which observed direct and synthetic pricing paths fail to align at a common information time.

The wedge-versus-arbitrage distinction can be written explicitly once bid and ask quotes are observed. Let \(b_t^j\) and \(a_t^j\) denote the best bid and ask for each leg
\[
j\in\{\pair{ETH/USD},\pair{ETH/BTC},\pair{BTC/USD}\},
\]
and define the log-midquote \(m_t^j=\{\log a_t^j+\log b_t^j\}/2\) and log half-spread \(h_t^j=\{\log a_t^j-\log b_t^j\}/2\). Let the midquote wedge be \(\epsilon_t=m_t^{\pair{ETH/USD}}-m_t^{\pair{ETH/BTC}}-m_t^{\pair{BTC/USD}}\), and let \(f\) be the per-leg taker fee.

\begin{proposition}[Midquote wedge and executable band]\label{prop:executable-band}
Under synchronized quotes, the absence of positive after-fee triangular round trips in both directions is equivalent to
\begin{equation}
-C_t \leq \epsilon_t \leq C_t,
\qquad
C_t \equiv h_t^{\pair{ETH/USD}}+h_t^{\pair{ETH/BTC}}+h_t^{\pair{BTC/USD}}-3\log(1-f).
\label{eq:executable-band}
\end{equation}
The forward executable return is \(\epsilon_t-C_t\), and the reverse executable return is \(-\epsilon_t-C_t\).
\end{proposition}

Proposition~\ref{prop:executable-band} clarifies the interpretation used throughout the paper. A nonzero midquote wedge is evidence of a parity deviation, but a positive executable arbitrage requires the wedge to cross the bid--ask plus fee band. The Gate.io extension below computes the forward and reverse round trips directly from observed bids and asks, so the reported breach count does not rely on a close-price or OHLC proxy.

This interpretation is especially important for Ethereum. If USD is treated as the only natural valuation anchor, one may be tempted to read deviations in the direct dollar price as absolute valuation errors. The triangular framework argues against that move. In a multi-numeraire environment, a useful prior question is whether Ethereum prices cohere across numeraires. The object of interest is therefore a no-arbitrage-style restriction in relative prices, not an absolute valuation benchmark.

\subsection{Baseline restriction versus extensions}
In the completed baseline evidence, the wedge is constructed from hourly Coinbase spot prices using an intersected, single-venue, true-USD triangle. This design supports a narrow interpretation of \(\epsilon_t\): parity deviations within one venue, absent stablecoin-dollar contamination. Three extensions add discipline to that interpretation. First, a Binance triangle with USDT legs yields a smaller unconditional wedge on average than the Coinbase baseline, which disciplines how the stablecoin channel should be interpreted. Second, a Gate.io order-book extension over the three stress windows constructs a genuine bid--ask executable band on a secondary USDT venue and finds no positive after-fee triangular round trip. Third, a proxy-midquote exercise based on hourly high--low midranges does not provide a valid historical top-of-book robustness check for the Coinbase baseline and is therefore not treated as core evidence. The paper therefore treats the Coinbase wedge as a reduced-form measure of relative pricing tension, while using the Gate exercise to show that the stress-window wedges need not imply executable arbitrage once spreads and fees are imposed.

\section{Data and Variable Construction}\label{sec:data}
\subsection{Baseline sample}
The baseline uses Coinbase spot data at hourly frequency from 2022-01-01 00:00:00 UTC through 2023-06-30 23:00:00 UTC. The three required markets are \pair{ETH/USD}, \pair{ETH/BTC}, and \pair{BTC/USD}. Synchronization is intersection only, with no forward fill. The resulting baseline contains 13{,}100 hourly observations.

This sample design is intentionally conservative. All three legs are native Coinbase spot markets. No leg is constructed from a stablecoin proxy for USD, and no cross-venue matching is required. As a result, the baseline evidence is insulated from several interpretation problems that would arise immediately in a broader sample, including reserve-risk contamination in a stablecoin-based dollar leg and pure cross-exchange settlement frictions.

\subsection{Sampling frequency}
The hourly sampling frequency is a baseline compromise rather than a claim that one hour is intrinsically optimal. Daily sampling is too coarse for a market that trades continuously and for stress episodes whose effects often occur within the day. It also imposes an arbitrary UTC close on a 24/7 market. At the other extreme, minute-level sampling would be attractive only with synchronized trades or historical top-of-book quotes for all three Coinbase legs. In their absence, a minute close-price triangle would be more exposed to asynchronous trades, bid--ask bounce, and microstructure noise than the hourly baseline.

To check that the main facts are not mechanical artifacts of the one-hour grid, I recompute the Coinbase true-USD wedge after coarsening the three price legs to four-hour and daily UTC bars, always using the last observed leg price in each bar and then recomputing the triangular wedge. Table~\ref{tab:frequency-choice} reports the result. The signed mean remains economically close to zero at all three frequencies, ADF diagnostics reject a unit root in every case, and the pooled stress-window mean absolute wedge remains larger than the calm-period mean. The daily design has only 22 event observations, so it is not used for event-by-event inference, but it is useful as a coarsening check. The hourly grid is therefore retained because it preserves intraday stress variation while avoiding the stronger synchronization assumptions that a full-sample minute design would require.

\begin{table}[htbp]
\centering
\caption{Sampling-frequency robustness in the Coinbase true-USD baseline}
\label{tab:frequency-choice}
\scriptsize
\begin{tabular}{@{}lrrrrrrr@{}}
\toprule
Frequency & $n$ & Mean $\epsilon$ & Mean $|\epsilon|$ & AR(1) & ADF & Event $|\epsilon|$ & Calm $|\epsilon|$ \\
\midrule
1 hour & 13{,}100 & -0.065 & 2.655 & 0.032 & -110.893 & 3.882 & 2.604 \\
4 hours & 3{,}276 & -0.088 & 2.673 & 0.064 & -53.655 & 3.526 & 2.637 \\
1 day & 546 & -0.043 & 2.649 & -0.002 & -23.344 & 4.520 & 2.570 \\
\bottomrule
\end{tabular}
\begin{minipage}{0.92\linewidth}
\vspace{0.5em}
\footnotesize \textit{Notes:} All wedge magnitudes are in basis points except AR(1) and ADF. Coarser frequencies are constructed by taking the last available price for each leg within each UTC bar and recomputing \(\epsilon_t\). Event observations pool the Terra/Luna, FTX, and USDC--SVB windows; daily event inference is descriptive because the pooled event sample contains only 22 daily observations.
\end{minipage}
\end{table}

\subsection{Wedge construction}
For each synchronized timestamp, the baseline wedge is constructed as in equation \eqref{eq:epsilon}. The baseline implementation uses hourly close prices. That choice is sufficient for the single-venue true-USD baseline, but it is not the most stringent microstructure design. For that reason, the paper distinguishes three objects: the baseline close-price wedge, an unsuccessful Coinbase high--low midrange proxy that is treated only as a data limitation, and a separate Gate.io order-book extension that uses actual bid and ask snapshots in stress windows.

To keep interpretation conservative, the paper uses both signed and absolute versions of the wedge. The signed wedge \(\epsilon_t\) is informative for average consistency and persistence diagnostics. The absolute wedge \(|\epsilon_t|\) is informative about the magnitude of parity deviations, especially during stress windows where widening matters more than sign.

\subsection{Gate.io order-book extension sample}
The executable-band extension uses Gate.io historical spot order-book files for \pair{ETH/USDT}, \pair{ETH/BTC}, and \pair{BTC/USDT} during the Terra/Luna, FTX, and USDC--SVB windows. For each hourly archive file, the construction uses only the initial full-depth snapshot block and extracts the best bid, best ask, midpoint, and half-spread for each leg. This produces one quote-based observation per market-hour. The design is intentionally snapshot-only: attempted replay of subsequent update rows generated crossed books under plausible update semantics, so the paper does not impose an undocumented order-book cleaning rule.

Across the three event windows, the archive contains 524 usable common hours out of 528 expected hours. The missing hours are two FTX hours on 2022-11-14 and two USDC--SVB hours on 2023-03-10; in each case all three Gate legs are missing for the same hour. They are dropped by common-hour intersection, with no interpolation and no cross-venue patching.

\subsection{State variables in the baseline}
The baseline state-dependent specification for \(|\epsilon_t|\) uses three right-hand-side variables: a 24-hour realized volatility measure for \pair{BTC/USD}, \(RV^{BTC}_{t,24h}\); the log of hourly Coinbase \pair{ETH/USD} volume; and an indicator, \(Event_t\), that equals one during the pre-specified Terra/Luna, FTX, and USDC--SVB stress windows.
These variables are best viewed as reduced-form state controls rather than structural mechanisms. In particular, the volume proxy is informative but still crude, and the baseline does not include fee, spread, funding, or order-book depth measures.

\section{Empirical Design}\label{sec:design}
\subsection{Completed baseline evidence}
The empirical design distinguishes sharply between headline baseline evidence and supporting extensions.

The completed baseline evidence consists of four elements. First, the paper summarizes the unconditional behavior of \(\epsilon_t\), including its mean, dispersion, and tails. Second, it evaluates unit-root behavior and short-run persistence using the reported ADF, KPSS, and AR(1) diagnostics. Third, it estimates a simple state-dependent regression for \(|\epsilon_t|\) using BTC volatility, \pair{ETH/USD} volume, and a pooled event indicator. Fourth, it compares wedge magnitudes during three stress episodes---Terra/Luna, FTX, and USDC--SVB---with calm periods.

In addition, three extensions are now sufficiently developed to be discussed in the main text with appropriately narrow interpretation. The Binance USDT extension contributes descriptive evidence on unconditional wedge magnitudes under a stablecoin-based dollar leg. The Gate.io order-book extension contributes a stress-window executable-band check using actual bid and ask snapshots on a secondary USDT venue. The Coinbase midquote extension contributes a methodological negative result: because no historical top-of-book midquote archive was available for the true-USD Coinbase triangle, an OHLC high--low midrange proxy could be constructed, but that object is not treated as a valid replacement for synchronized historical midquotes in triangular parity testing.

\subsection{Baseline estimating equations}
The unit-root and persistence diagnostics can be summarized with the following reduced-form law of motion:
\begin{equation}
\epsilon_t = c + \phi \epsilon_{t-1} + u_t,
\label{eq:ar1}
\end{equation}
where a small \(|\phi|\) together with ADF rejection of a unit root indicates that parity deviations do not drift persistently in the baseline sample.

The state-dependent specification is
\begin{equation}
|\epsilon_t|_{\bp} = a + b_1 RV^{BTC}_{t,24h} + b_2 \log Volume^{\pair{ETH/USD}}_t + b_3 Event_t + u_t,
\label{eq:state}
\end{equation}
where \(|\epsilon_t|_{\bp}\) is the absolute wedge measured in basis points and standard errors are Newey--West HAC \citep{newey_west_1987}.

Equation \eqref{eq:state} is used cautiously. It is informative about state dependence, but it does not separately identify liquidity effects, inventory constraints, transaction-cost bands, or stablecoin-dollar stress.

\subsection{Scope of extensions}
The evidence has different inferential weight, so the paper separates the headline baseline from supporting tests. The Gate.io stress-window extension is the only completed quote-based test of executable triangular returns, and it is used only for the available event-hour snapshots. Re-estimating the wedge at one- to five-minute frequency would still be useful, because it would help distinguish genuine short-run dislocations from hourly sampling effects. The Binance USDT extension is informative about unconditional wedge magnitudes and now enters the stress-interaction design, but it remains a supporting comparison rather than a replacement for the Coinbase true-USD baseline. The longer 2019--2026 Coinbase and Binance panels are also useful, but mainly as appendix context: they confirm near-zero means and low persistence while spanning market regimes that differ from the pre-specified 2022--2023 event design.

Cross-venue and long-sample evidence therefore discipline the interpretation without expanding the headline claim beyond the Coinbase true-USD baseline. Table~\ref{tab:evidence-hierarchy} summarizes this hierarchy.

\begin{table}[htbp]
\centering
\caption{Evidence hierarchy}
\label{tab:evidence-hierarchy}
\small
\begin{tabular}{@{}>{\raggedright\arraybackslash}p{0.20\linewidth}>{\raggedright\arraybackslash}p{0.29\linewidth}>{\raggedright\arraybackslash}p{0.41\linewidth}@{}}
\toprule
Status & Evidence & Role in interpretation \\
\midrule
Headline baseline & Coinbase true-USD hourly triangle, 2022-01-01 to 2023-06-30 & Defines the paper's main empirical object: a single-venue, true-USD triangular wedge with intersection-only synchronization. \\
Supporting extension & Binance USDT triangle and Gate.io stress-window bid--ask snapshots & Disciplines stablecoin and executable-arbitrage interpretations, but remains subordinate to the baseline. \\
Appendix robustness & Event-window sensitivity, pseudo-triangle placebo, structural-break and long-sample summaries & Checks whether the baseline facts are fragile without adding new headline claims. \\
Future agenda & Coinbase true-USD executable band, higher-frequency quote reconstruction, broader matched venue panels & Requires additional data or stronger synchronization discipline before it can support publishable main claims. \\
\bottomrule
\end{tabular}
\begin{minipage}{0.92\linewidth}
\vspace{0.5em}
\footnotesize \textit{Notes:} The table is a scope-control device. It prevents the supporting Binance, Gate, and long-sample evidence from being read as replacing the narrow Coinbase baseline or as establishing market-wide no-arbitrage.
\end{minipage}
\end{table}

\section{Baseline Results}\label{sec:baseline}
\subsection{Unconditional properties of the wedge}
The first baseline result is that the Coinbase true-USD triangle satisfies the pricing restriction closely on average. The signed wedge has a sample mean of \(-6.461 \times 10^{-6}\) log points, or roughly \(-0.065\bp\), over 13{,}100 hourly observations. Its standard deviation is 3.868\bp, its mean absolute value is 2.655\bp, its 5th to 95th percentile range is from \(-5.83\bp\) to \(+5.68\bp\), and its extrema are approximately \(-35.14\bp\) and \(+41.50\bp\).

\begin{figure}[htbp]
\centering
\includegraphics[width=0.88\linewidth]{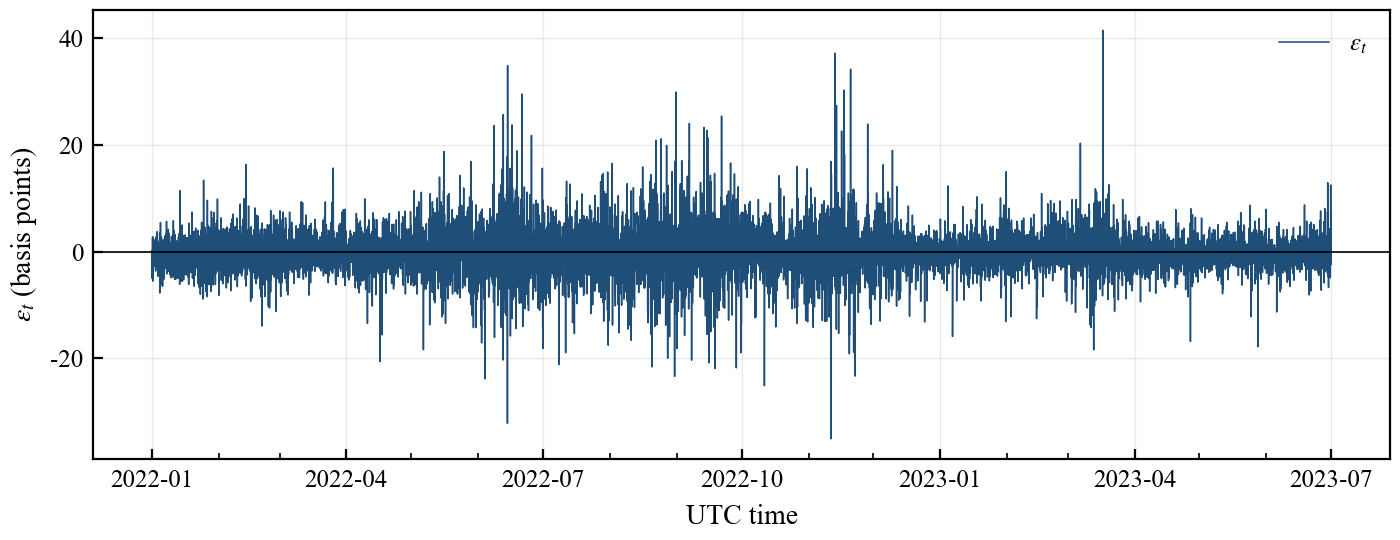}
\caption{Coinbase true-USD wedge time series. The figure plots the hourly triangular-pricing wedge from 2022-01-01 through 2023-06-30 using UTC timestamps and intersection-only synchronization. The y-axis reports \(\epsilon_t\) in basis points, where \(1\bp=10^{-4}\) log units.}
\label{fig:coinbase-series}
\end{figure}

\begin{figure}[htbp]
\centering
\includegraphics[width=0.78\linewidth]{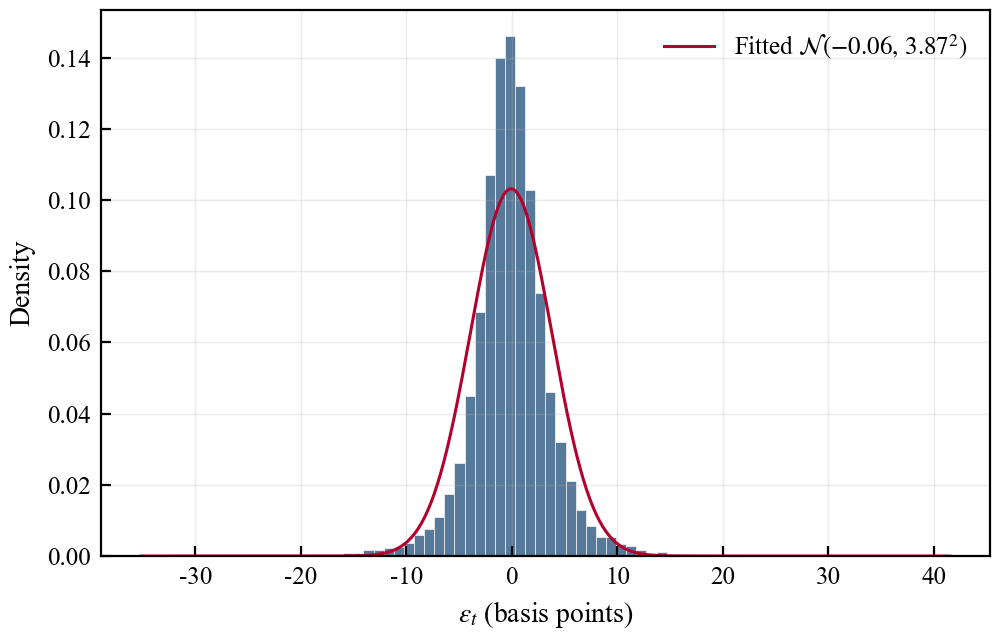}
\caption{Coinbase true-USD wedge distribution. The histogram uses the same hourly Coinbase true-USD baseline sample as Figure~\ref{fig:coinbase-series}. The x-axis reports \(\epsilon_t\) in basis points; the fitted normal curve is included only as a visual benchmark.}
\label{fig:coinbase-hist}
\end{figure}

Taken together, these moments suggest a clear baseline fact: within a single venue and with true USD legs, direct and synthetic \pair{ETH/USD} prices line up tightly on average, even though economically nontrivial deviations occur in the tails. This is exactly the pattern one would expect if triangular consistency is a useful first-order restriction but not a frictionless identity at every instant. Table~\ref{tab:coinbase-summary} reports the baseline summary statistics, while Figures~\ref{fig:coinbase-series} and \ref{fig:coinbase-hist} show the full hourly series and unconditional distribution.

\begin{table}[htbp]
\centering
\caption{Coinbase true-USD baseline summary statistics}
\label{tab:coinbase-summary}
\begingroup
\scriptsize
\setlength{\tabcolsep}{3pt}
\begin{tabular}{@{}lrrrrrrrr@{}}
\toprule
Variable & $n$ & Mean & Std. Dev. & Min & P5 & Median & P95 & Max \\
\midrule
$\epsilon_t$ & 13{,}100 & $-6.461\times10^{-6}$ & $3.868\times10^{-4}$ & $-0.003514$ & $-0.000583$ & $-1.269\times10^{-5}$ & $0.000568$ & $0.004150$ \\
$|\epsilon_t|_{\bp}$ & 13{,}100 & 2.655 & 2.814 & 0.00002 & 0.167 & 1.874 & 7.744 & 41.502 \\
\bottomrule
\end{tabular}
\begin{minipage}{0.92\linewidth}
\vspace{0.5em}
\footnotesize \textit{Notes:} Coinbase spot triangle with \pair{ETH/USD}, \pair{ETH/BTC}, and \pair{BTC/USD}, sampled hourly from 2022-01-01 through 2023-06-30 using intersection-only synchronization and no forward fill. The signed wedge is in log units; \(|\epsilon_t|_{\bp}\) is the absolute wedge in basis points.
\end{minipage}
\endgroup
\end{table}

\subsection{Unit-root diagnostics and mean reversion}
The second baseline result is that the wedge does not behave like a slowly drifting gap. The ADF statistic is \(-110.893\), well beyond conventional critical values, rejecting the null of a unit root. Lag-augmented ADF checks through 168 hourly lags also reject the unit-root null. The AR(1) coefficient is approximately \(\phi=0.0315\), with an implied half-life of about 0.20 hours.

These estimates matter for how the object should be interpreted. A low-persistence wedge is not consistent with a slowly moving intrinsic-value gap. It is more consistent with a relative pricing discrepancy that is usually corrected quickly, albeit not instantaneously, in a market with small but nonzero frictions. A KPSS level-stationarity diagnostic rejects strict unconditional level stationarity, so the safer interpretation is not that the process is invariant across states. Rather, the signed wedge has a near-zero mean and low persistence, while its dispersion changes with stress conditions. The low first-order autocorrelation and rapid decay of the autocorrelation function are aligned with that interpretation.

\subsection{Binance USDT extension}
\begin{figure}[htbp]
\centering
\includegraphics[width=0.88\linewidth]{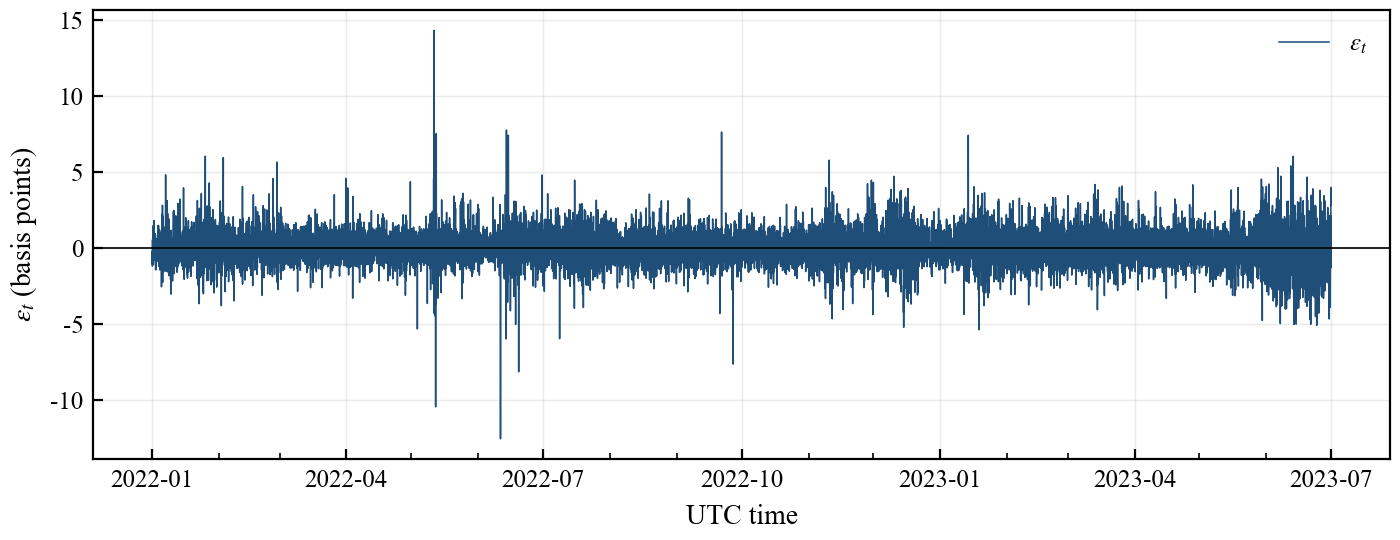}
\caption{Binance USDT wedge time series. The figure plots the hourly triangular-pricing wedge from 2022-01-01 through 2023-06-30 using UTC timestamps and intersection-only synchronization. The y-axis reports \(\epsilon_t\) in basis points, where \(1\bp=10^{-4}\) log units.}
\label{fig:usdt-series}
\end{figure}

\begin{figure}[htbp]
\centering
\includegraphics[width=0.78\linewidth]{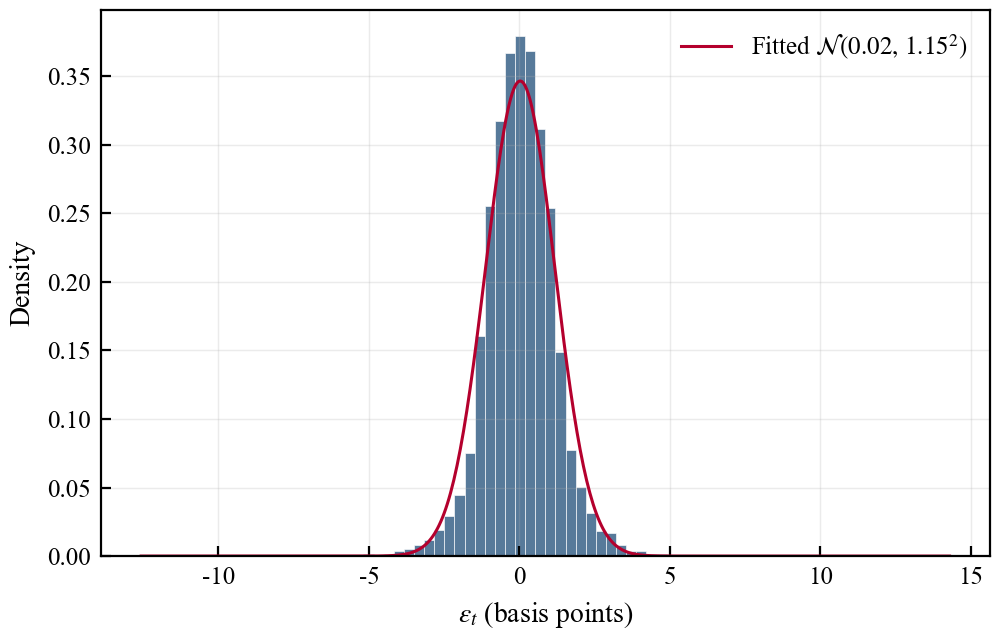}
\caption{Binance USDT wedge distribution. The histogram uses the same hourly Binance USDT extension sample as Figure~\ref{fig:usdt-series}. The x-axis reports \(\epsilon_t\) in basis points; the fitted normal curve is included only as a visual benchmark.}
\label{fig:usdt-hist}
\end{figure}

The completed Binance extension uses the same triangular restriction over the same calendar window, but with Binance spot pairs \pair{ETH/USDT}, \pair{ETH/BTC}, and \pair{BTC/USDT}. Its main contribution is descriptive rather than causal. The average wedge is tighter than in the Coinbase true-USD baseline: the mean absolute wedge is 0.862\bp, compared with 2.655\bp{} in the Coinbase sample, and the standard deviation falls from 3.868\bp{} to 1.152\bp. The signed mean remains close to zero at \(+0.024\bp\). Table~\ref{tab:usdt-summary} and Figures~\ref{fig:usdt-series}--\ref{fig:usdt-hist} summarize these unconditional properties.

This comparison is informative because it runs against a simple prior that stablecoin denomination might mechanically produce a looser wedge on average. In this sample, it does not. The conservative inference is therefore that venue liquidity and arbitrage efficiency are important alongside numeraire type for unconditional wedge magnitudes. That does not eliminate stablecoin-related mechanisms. Rather, it narrows their interpretation: absent sharper stress-state evidence, stablecoin effects are better framed as possible tail-event mechanisms than as the dominant unconditional driver of parity deviations.

The USDT extension remains deliberately limited. It is not a second headline baseline and should not be read as a cross-venue ranking of pricing quality. Its event-state role is addressed only through the reduced-form stress-interaction regression below, which shows that the Binance USDT response is smaller than the Coinbase response in the FTX and USDC--SVB windows. That evidence narrows the stablecoin interpretation but does not identify a causal mechanism.

\begin{table}[htbp]
\centering
\caption{Binance USDT extension summary statistics}
\label{tab:usdt-summary}
\begingroup
\scriptsize
\setlength{\tabcolsep}{3pt}
\begin{tabular}{@{}lrrrrrrrr@{}}
\toprule
Variable & $n$ & Mean & Std. Dev. & Min & P5 & Median & P95 & Max \\
\midrule
$\epsilon_t$ & 13{,}103 & $2.407\times10^{-6}$ & $1.152\times10^{-4}$ & $-0.001257$ & $-0.000170$ & $2.279\times10^{-6}$ & $0.000180$ & $0.001429$ \\
$|\epsilon_t|_{\bp}$ & 13{,}103 & 0.862 & 0.764 & 0.000002 & 0.064 & 0.699 & 2.272 & 14.288 \\
\bottomrule
\end{tabular}
\begin{minipage}{0.92\linewidth}
\vspace{0.5em}
\footnotesize \textit{Notes:} Binance spot triangle with \pair{ETH/USDT}, \pair{ETH/BTC}, and \pair{BTC/USDT}, sampled hourly from 2022-01-01 through 2023-06-30 using intersection-only synchronization and no forward fill. The signed wedge is in log units; \(|\epsilon_t|_{\bp}\) is the absolute wedge in basis points.
\end{minipage}
\endgroup
\end{table}

\subsection{Midquote extension as a methodological limitation}
The attempted midquote extension does not alter the paper's main evidence. No historical top-of-book bid--ask archive is available for the full Coinbase true-USD sample. Coinbase's public historical endpoint provides OHLCV candles rather than synchronized historical best bid and ask quotes, and the available ticker endpoint gives only a contemporaneous snapshot. The extension therefore used an hourly high--low midrange proxy, \((H_t+L_t)/2\), for each leg.

That proxy is not a valid substitute for synchronized historical midquotes in a triangular pricing test. It is a midpoint over realized intrabar trade prices, not a common-time quote midpoint, and it can therefore import aliasing and intrabar range effects that are conceptually distinct from a quote-based wedge. Consistent with that concern, the proxy-midrange wedge is substantially wider and has very low correlation with the close-based baseline series. The paper accordingly presents the exercise as a methodological negative result: it shows why true historical top-of-book quote data are needed for a publishable midquote or executable-band extension, not that the baseline findings are robust to a quote-based redesign.

\subsection{Gate.io executable-band extension}
The Gate.io order-book extension addresses the same concern from a different angle. Rather than approximating Coinbase midquotes with OHLC data, it uses historical Gate order-book snapshots for \pair{ETH/USDT}, \pair{ETH/BTC}, and \pair{BTC/USDT} in the three stress windows. The sample contains 524 common event-hour observations after dropping four missing common hours from the official archive. Each observation is built from the initial full-depth hourly snapshot, yielding synchronized best bid, best ask, midquote, and half-spread measures for all three legs.

The midquote wedge remains small in this quote-based stress-window sample. Across all Gate event hours, the mean wedge is \(-0.172\bp\), the mean absolute wedge is 2.660\bp, and the maximum absolute wedge is 21.435\bp. More importantly for interpretation, the average three-leg spread-plus-fee band is 36.825\bp, consisting of 6.810\bp{} in summed half-spreads and 30.015\bp{} in taker-fee cost under the 10 bp per-leg taker-fee assumption, which matches the public VIP0 spot taker schedule on Gate's fee page \citep{gate_fee_schedule}. The largest net executable gap is \(-23.102\bp\), so even the closest event-hour remains inside the executable band by more than 23 bp. The breach count is zero in every event. A sensitivity check with a lower 5 bp per-leg fee also produces zero breaches, with the closest reverse-direction gap still negative at \(-8.091\bp\). Table~\ref{tab:gate-executable} summarizes the primary 10 bp specification.

\begin{table}[htbp]
\centering
\caption{Gate.io order-book executable-band extension}
\label{tab:gate-executable}
\scriptsize
\begin{tabular}{@{}lrrrrrrrr@{}}
\toprule
Event & $n$ & Mean $|\epsilon|$ & Spread & Fee & Spread+fee & Max gap & Min cushion & Breaches \\
\midrule
Terra/Luna & 216 & 2.925 & 9.785 & 30.015 & 39.800 & -23.102 & 23.102 & 0 \\
FTX & 214 & 2.561 & 5.942 & 30.015 & 35.957 & -24.170 & 24.170 & 0 \\
USDC--SVB & 94 & 2.278 & 1.949 & 30.015 & 31.964 & -24.119 & 24.119 & 0 \\
All & 524 & 2.660 & 6.810 & 30.015 & 36.825 & -23.102 & 23.102 & 0 \\
\bottomrule
\end{tabular}
\begin{minipage}{0.92\linewidth}
\vspace{0.5em}
\footnotesize \textit{Notes:} All non-count entries are in basis points. Gate.io spot order-book triangle with \pair{ETH/USDT}, \pair{ETH/BTC}, and \pair{BTC/USDT}, sampled once per available event-hour from the initial full-depth snapshot of each hourly archive file. Spread is the sum of the three leg-level half-spreads. Fee is \(-3\log(1-f)\) for \(f=0.001\). Spread+fee is the approximate executable band width in Proposition~\ref{prop:executable-band}. Max gap is the larger of the two actual after-fee executable round-trip returns within each event. Min cushion is the smallest value of \(-\max\{g^{fwd}_t,g^{rev}_t\}\); positive values indicate that the nearest observed round trip remains inside the executable band. A breach is a positive net executable gap in either direction. The diagnostics file also reports 5, 10, 15, and 20 bp per-leg fee sensitivities.
\end{minipage}
\end{table}

This result does not replace the Coinbase true-USD baseline, because the Gate triangle is a secondary USDT venue and the sampling rule is hourly snapshot-only. It does, however, directly address the main bid--ask objection in the data that are available: in the stress-window order-book snapshots, the measured quote wedges remain inside the stated spread-and-fee benchmark rather than becoming positive after-fee triangular returns.

\subsection{State dependence in the baseline sample}
The third baseline result is that parity deviations widen in states associated with stress. To separate generic stress from a stablecoin-venue interpretation, the stress-interaction specification stacks the Coinbase true-USD triangle with the Binance USDT supporting extension over common hourly timestamps. The omitted category is Coinbase during calm hours. The event coefficients therefore measure Coinbase true-USD event shifts, while the USDT interaction terms measure whether the Binance USDT response differs from the Coinbase response in the same calendar events.

The pattern is useful for interpretation. Conditional on BTC realized volatility and ETH quote-volume controls, the FTX and USDC--SVB coefficients are positive in the Coinbase true-USD baseline. The corresponding USDT interaction terms are negative and statistically significant, meaning that the Binance USDT wedge response is smaller, not larger, in those stress windows. Table~\ref{tab:state-reg} reports the specification. The result does not identify a causal mechanism, but it disciplines the stablecoin interpretation: in the completed data, stablecoin denomination is not supported as the dominant unconditional or event-state explanation for wedge magnitudes.

\begin{table}[htbp]
\centering
\caption{Stress-state interaction regression for wedge magnitude}
\label{tab:state-reg}
\small
\begin{tabular}{lrrrr}
\toprule
Variable & Coef. & NW S.E. & $t$ & $p$ \\
\midrule
Constant & -0.1980 & 0.1815 & -1.0909 & $\geq 0.10$ \\
$RV^{BTC}_{t,24h}$ & 0.0008 & 0.0002 & 4.7656 & $<0.001$ \\
$\log Volume^{ETH}_t$ & 0.2951 & 0.0235 & 12.5421 & $<0.001$ \\
Binance USDT & -2.0477 & 0.0529 & -38.6928 & $<0.001$ \\
Terra/Luna & 0.2033 & 0.2384 & 0.8529 & $\geq 0.10$ \\
FTX & 1.3779 & 0.4310 & 3.1968 & $<0.01$ \\
USDC--SVB & 1.3549 & 0.5816 & 2.3297 & $<0.05$ \\
USDT $\times$ Terra/Luna & -0.4361 & 0.2483 & -1.7568 & $<0.10$ \\
USDT $\times$ FTX & -1.4405 & 0.4410 & -3.2663 & $<0.01$ \\
USDT $\times$ USDC--SVB & -1.5358 & 0.5827 & -2.6354 & $<0.01$ \\
\midrule
$n$ & \multicolumn{4}{r}{26{,}150} \\
$R^2$ & \multicolumn{4}{r}{0.1866} \\
NW lag & \multicolumn{4}{r}{29} \\
\bottomrule
\end{tabular}
\begin{minipage}{0.92\linewidth}
\vspace{0.5em}
\footnotesize \textit{Notes:} Dependent variable is \(|\epsilon_t|\) in basis points. The panel stacks Coinbase true-USD and Binance USDT triangles on common hourly timestamps from 2022-01-01 through 2023-06-30. Coinbase calm hours are the omitted category. Event coefficients are Coinbase event shifts; USDT interaction terms measure the differential Binance USDT response. Newey--West HAC standard errors use Bartlett weights with \(L=\lfloor n^{1/3}\rfloor\). The design is reduced-form and should not be read as causal mechanism evidence.
\end{minipage}
\end{table}

\section{Stress Episodes}\label{sec:stress}
\begin{figure}[htbp]
\centering
\includegraphics[width=0.88\linewidth]{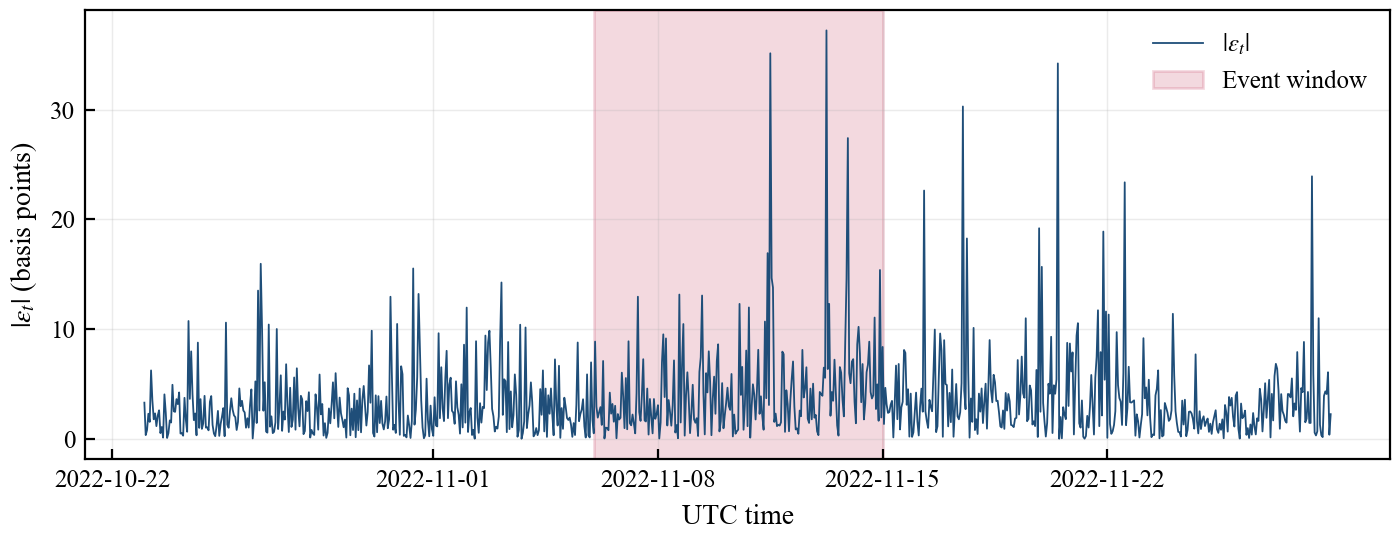}
\caption{FTX stress window in the Coinbase true-USD baseline. The figure plots \(|\epsilon_t|\) in basis points for the hourly Coinbase true-USD triangle around the 2022-11-06 to 2022-11-14 UTC event window. The shaded region marks the event window; the displayed sample includes a 14-day buffer on each side.}
\label{fig:ftx-event}
\end{figure}

The paper examines three stress windows in the same Coinbase baseline: Terra/Luna in May 2022, the FTX collapse in November 2022, and the USDC--SVB weekend in March 2023. In each case, the mean absolute wedge is higher than in the calm baseline. The mean absolute wedge rises from 2.604\bp{} in calm periods to 3.161\bp{} during Terra/Luna, 4.472\bp{} during FTX, and 4.176\bp{} during the USDC--SVB window. The strongest baseline event evidence is associated with the FTX episode, which has the largest Welch statistic in absolute value. Table~\ref{tab:event-study} summarizes the event windows, and Figure~\ref{fig:ftx-event} plots the FTX episode directly.

These patterns support a restrained conclusion: the triangular wedge widens during stress. They do not, on their own, establish why. In the baseline design, one defensible interpretation is that stress raises frictions, slows cross-rate adjustment, changes balance-sheet capacity, or induces short-run risk repricing across the ETH and BTC legs. Because the baseline triangle does not use USDT or USDC as the dollar leg, the stress results cannot be taken as evidence that stablecoin-specific dollar-proxy mechanisms are driving the wedge in this sample.

\begin{table}[htbp]
\centering
\caption{Coinbase baseline event-study summary}
\label{tab:event-study}
\begingroup
\small
\setlength{\tabcolsep}{3pt}
\begin{tabular}{@{}lrrrrr@{}}
\toprule
Event & $n$ & Mean $|\epsilon_t|$ & SD & Outside & Welch $t$ \\
\midrule
Terra/Luna & 216 & 3.161 & 4.221 & 8.80\% & 2.886 \\
FTX & 216 & 4.472 & 6.602 & 15.28\% & 5.644 \\
USDC--SVB & 96 & 4.176 & 5.561 & 16.67\% & 4.142 \\
Calm baseline & 12{,}572 & 2.604 & 3.782 & 4.96\% & 0.000 \\
\bottomrule
\end{tabular}
\endgroup
\begin{minipage}{0.92\linewidth}
\vspace{0.5em}
\footnotesize \textit{Notes:} Mean and SD are reported in basis points. The outside column reports the share of observations with \(|\epsilon_t|\) above twice the calm-period standard deviation of \(|\epsilon_t|\). Event windows are Terra/Luna 2022-05-07--2022-05-15, FTX 2022-11-06--2022-11-14, and USDC--SVB 2023-03-10--2023-03-13. Event-window boundaries are fixed from the episode dates rather than estimated from the wedge series.
\end{minipage}
\end{table}

\section{Discussion}\label{sec:discussion}
The discussion follows directly from the scope of the baseline evidence. The paper's first claim is narrow: Ethereum pricing can be productively analyzed as a relative asset-pricing problem across numeraires. That claim rests on the triangular restriction itself and on the empirical fact that the wedge is near zero on average in the narrowest baseline sample. The wedge is therefore informative precisely because the identity usually holds closely, not because it fails dramatically all the time.

The second claim is also narrow: when the wedge widens, the baseline evidence is more consistent with frictions and stress-period risk repricing than with persistent intrinsic mispricing. Three features point in that direction. First, ADF diagnostics reject a unit root and AR(1) persistence is low. Second, wedge magnitudes are larger during market stress windows. Third, the baseline excludes USDT and related stablecoin-dollar legs, so the observed widening cannot be mechanically attributed to dollar-proxy instability in this sample.

A placebo falsification reinforces that interpretation. If the measured wedge were merely a generic artifact of stitching together three log-price series, then breaking the contemporaneous identity should produce a similar residual. It does not. A pseudo-wedge that lags the \pair{ETH/BTC} leg by 24 hours has mean absolute magnitude of 132.35\bp, compared with 2.66\bp{} for the true contemporaneous wedge on the matched Coinbase sample, and its AR(1) coefficient is 0.958 rather than 0.032. The wedge behavior documented in the baseline is therefore a property of contemporaneous triangular pricing consistency, not a generic algebraic residual.

The newly completed Binance USDT extension further disciplines the interpretation. In unconditional average terms, the USDT triangle is tighter, not looser, than the Coinbase true-USD baseline. That fact does not show that stablecoin denomination is irrelevant. It suggests that one should not treat stablecoin status as the dominant unconditional explanation for measured wedge size in this sample. A more defensible reading is that venue liquidity, trading depth, and arbitrage efficiency matter in the unconditional average, while stablecoin-related mechanisms remain plausible candidates for tail behavior in stressed states.

The Gate.io order-book extension adds a different kind of discipline. It shows that, in the available stress-window order-book snapshots, measured midquote wedges do not translate into positive after-fee executable triangular returns. This strengthens the paper's distinction between a relative pricing wedge and an arbitrage profit. A nonzero \(\epsilon_t\) can be economically informative about market stress without being directly monetizable once the bid--ask band and taker fees are imposed.

The Coinbase midquote exercise points in a different direction. It does not strengthen the evidence base; rather, it clarifies a data limitation. Without a historical top-of-book bid--ask archive for the Coinbase true-USD triangle, the high--low midrange proxy cannot be read as a genuine midquote robustness check. The paper therefore refrains from drawing substantive inference from that exercise and treats the Gate analysis as the only completed quote-based executable-band extension.

That interpretation remains intentionally bounded. The paper does not claim to separate market microstructure frictions from broader risk-bearing constraints. It does not construct a Coinbase true-USD executable band, and the Gate extension is hourly snapshot-only rather than a reconstructed one-minute order book. It also does not establish whether stablecoin-based dollar legs amplify tail behavior in matched event-state data. Those questions are natural extensions, but they are outside the headline baseline claim.

The long-sample outputs are useful in the same disciplined way. The 2019--2026 Coinbase and Binance panels preserve the near-zero mean and low-persistence pattern, but the long window spans materially different regimes, including the COVID aftermath and post-2023 market structure. For that reason, the long sample is best read as appendix context that supports scope, not as a replacement for the pre-specified 2022--2023 event design.

This separation between baseline findings and supporting extensions keeps the claim bounded. The paper has a defined relative pricing wedge, a conservative single-venue true-USD design, evidence of near-zero mean and low persistence, state-dependent widening during stress, and cross-venue, placebo, and executable-band checks that discipline overinterpretation. Further mechanism analysis can build on this structure without changing the interpretation of the baseline object.

\section{Conclusion}\label{sec:conclusion}
This paper advances a simple argument. Ethereum need not be approached only as a dollar-denominated absolute valuation problem; it can also be studied as a relative asset-pricing problem across numeraires. The empirical counterpart of that argument is a triangular pricing restriction linking \pair{ETH/USD}, \pair{ETH/BTC}, and \pair{BTC/USD}, and the central empirical object is the relative pricing wedge \(\epsilon_t\).

In the baseline evidence from a single-venue Coinbase true-USD triangle, that wedge has a near-zero mean, rejects a unit root in ADF diagnostics, and has low first-order persistence. The wedge also widens during stress episodes, with the strongest baseline episode around the FTX collapse. A completed Binance USDT extension shows that the average wedge can be even tighter under a stablecoin-based dollar proxy, which suggests that venue liquidity and arbitrage efficiency matter alongside numeraire type for unconditional wedge magnitudes. A Gate.io order-book extension further shows that the available stress-window bid--ask snapshots do not contain positive after-fee executable triangular breaches. Stablecoin-related interpretation is therefore best reserved for tail-event mechanisms to be tested more directly in future work, not treated as the dominant unconditional driver in this paper.

The paper's contribution is therefore best understood as an asset-pricing restriction paper. It provides a disciplined way to study Ethereum pricing across numeraires before committing to any particular absolute valuation model. The attempted Coinbase midquote extension underscores that this agenda requires genuine historical quote data rather than OHLC-based proxies, while the Gate extension shows how the same framework can be sharpened when bid and ask archives are available. Future work should extend the framework to higher-frequency quote reconstruction, broader matched venue panels, and true-USD executable bands to help separate generic market frictions from genuinely numeraire-specific stress.

\section*{Declarations}

\noindent\textit{Data and code availability.}
The analysis uses historical Coinbase and Binance spot data and Gate.io order-book archive data. Source code, variable documentation, canonical result files, and processed analysis inputs are available from the author upon reasonable request, subject to exchange terms and any applicable redistribution restrictions. Raw exchange archives are not redistributed where file size or exchange terms make redistribution impractical.

\noindent\textit{Funding.}
This research did not receive any specific grant from funding agencies in the public, commercial, or not-for-profit sectors.

\noindent\textit{Declaration of competing interest.}
The author declares no known competing financial interests or personal relationships that could have appeared to influence the work reported in this paper.

\noindent\textit{Author contribution.}
Hongzhe Wen: conceptualization, data curation, methodology, software, formal analysis, visualization, writing, and revision.

\noindent\textit{Declaration of generative AI and AI-assisted technologies in the writing process.}
AI-assisted tools, including OpenAI Codex and Anthropic Claude, were used for language editing, code-review support, and workflow organization. The author reviewed and verified the analysis, interpretation, citations, and final text and remains responsible for the content of the publication.

\end{document}